\documentstyle[aps,prl,psfig,floats]{revtex}
\begin{document}
\draft
\def\rhopar{\rho_{_\|}}
\def\etbr{$\kappa-$(ET)$_2$Cu[N(CN)$_2$]Br}
\def\ets{$\kappa-$(ET)$_2$Cu(NCS)$_2$}
\def\ybco{YBa$_2$Cu$_3$O$_{6.95}$}

\twocolumn[\hsize\textwidth\columnwidth\hsize\csname %
@twocolumnfalse\endcsname

\title{Low Temperature Penetration depth of \etbr \\ and 
$\kappa-$(ET)$_2$Cu(NCS)$_2$}
\author{A. Carrington}
\address{Department of Physics, University of Illinois at Urbana-Champaign, 
1110 West Green St, Urbana, 61801 Illinois \\ and Department of Physics and Astronomy, University of Leeds, Leeds LS2 9JT, England.} 
\author{I. J. Bonalde, R. Prozorov, R. W. Giannetta}
\address{Department of Physics, University of Illinois at Urbana-Champaign, 1110 West Green St, Urbana, 61801 Illinois}
\author{A. M. Kini, J. Schlueter, H. H. Wang, U. Geiser and J.M. Williams}
\address{Chemistry and Materials Science Division, Argonne National Laboratory, Argonne, Illinois 60439}
\date{\today}
\maketitle

\begin{abstract}
We present high precision measurements of the penetration depth of single crystals of $\kappa-$(ET)$_2$Cu\mbox{[}N(CN)$_2$\mbox{]}Br and \ets\ at temperatures down to 0.4~K.  We find that, at low temperatures, the in-plane penetration depth ($\lambda_\|$) varies as a fractional power law, $\lambda_\|\sim T^\frac{3}{2}$.  Whilst this may be taken as evidence for novel pair excitation processes, we show that the data are also consistent with a quasilinear variation of the superfluid density, as is expected for a $d$-wave superconductor with impurities or a small residual gap. Our data for the interplane penetration depth show similar features and give a direct measurement of the absolute value, $\lambda_\bot(0)=100\pm20$~$\mu$m.
\end{abstract}
]


Compounds of the family $\kappa-$(ET)$_2$X have the highest transition temperatures of all organic superconductors known to date\cite{williams90}.  They have recently attracted considerable attention because of their similarity to the high $T_c$ cuprates and the possibility that they may also have a non-conventional paring state \cite{mckenzie97}. The two materials studied here, \etbr\ ($T_c\sim$ 11.6~K) and \ets\ ($T_c \sim$ 9.6~K), are highly anisotropic, layered, extreme type II superconductors.  As in the cuprates, the superconducting phase in these materials is in close proximity to an antiferromagnetic phase. Both antiferromagnetic spin fluctuations and a pseudo-gap have been detected in NMR measurements in the normal state \cite{kawamoto95}.  Neither the underlying pairing mechanism nor the symmetry of the order parameter has been conclusively established.  Although NMR \cite{mayaffre95,desoto95}, specific heat \cite{nakazawa97} and thermal conductivity \cite{belin98} measurements all suggest a non-conventional pairing state, results of penetration depth measurements have been inconsistent, with evidence for both conventional \cite{harshman90,lang92} and non-conventional  \cite{kanoda90,le92,achkir93} behavior.  However, none of these penetration depth measurements have been performed over a temperature range ($T/T_c$) and a precision, comparable to those in the cuprates \cite{hardy93}.   In this Letter, we present measurements of both the in-plane $\lambda_\|$, and the interplane, $\lambda_\bot$, penetration depths in \etbr\ and \ets\ at temperatures down to 0.4~K.

Our measurements were performed on single crystals of \etbr\ and \ets\ which were grown at Argonne National Laboratory.  Details of the growth procedures have been given elsewhere \cite{kini90}.  Penetration depth measurements were performed using a 13~MHz tunnel diode oscillator  \cite{carrington99} mounted on  a $^3$He refrigerator.  The low noise level [$\frac{\Delta F}{F_0} \simeq 10^{-9}$], and low drift of the oscillator allows us to obtain high resolution data with a very small temperature spacing interval.  The samples were attached, with a small amount of vacuum grease,  to a sapphire rod which fitted inside the copper sense coil.  The sense coil was calibrated using spheres of Aluminum.  The sample temperature was measured with a calibrated Cernox thermometer attached to the other end of the sapphire rod.  The samples were cooled slowly (0.1-1.0 K/min) to avoid introducing disorder \cite{aburto98}.  The entire cryostat was surrounded by a triple mumetal shield that reduced the stray dc field to less than 0.005~Oe.  The RF probe field was estimated to be $\sim 0.005$~Oe.   This very low measurement field is important as in these materials $H_{c1}$ is very low (especially for $H$ parallel to the layers) and the vortices are very weakly pinned in the mixed state\cite{mansky94,tea98}.  The complex vortex dynamics are a serious problem for extracting $\lambda(T)$ from mixed state magnetization measurements \cite{lee97}, and have not been fully taken into account in some previous studies \cite{harshman90,lang92}.  

For measurements of $\lambda_\|$, the RF field was directed normal to the conducting planes thus ensuring that only in-plane supercurrents are excited \cite{averageinplane}.  We relate the measured frequency shifts to changes in $\lambda$ by measuring the area of the faces parallel to the field and multiplying by a demagnetization factor, which was estimated by the inscribed or exscribed ellipsoid approximation \cite{stoner45}.  The aspect ratios of our crystals varied from approximately 1.0 to 2.5 (the short dimension is normal to the planes), and so the demagnetizing correction $1/(1-N)$ was between 1.6 and 2.4.  We estimate that this calibration is accurate to $\sim$20\%, which is in accord with the measured variations of $\Delta\lambda_\|$ between samples.    To determine $\lambda_\bot$ the field was directed along the planes.  The strong anisotropy of these materials ensures that in this orientation the response is dominated by interlayer currents for our crystals' aspect ratios. 

\begin{figure}
\centerline{\psfig{figure=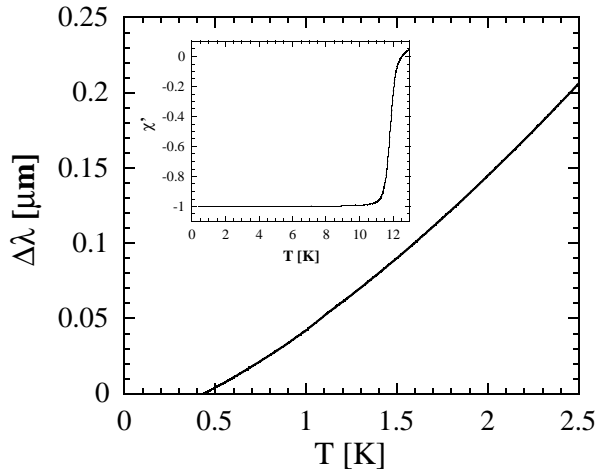,width=8cm}}
\vskip 0cm
\caption{Temperature dependence of the in-plane penetration depth $\lambda_\|$ of a crystal of \etbr.  The inset shows the susceptibility over the full temperature range.}
\label{lfig}
\end{figure}

\begin{figure}
\centerline{\psfig{figure=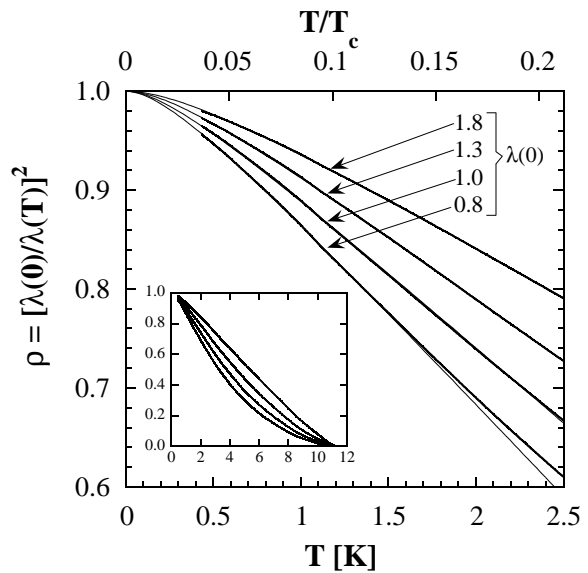,width=8cm}}
\vskip 0cm
\caption{In-plane superfluid density $\rhopar=\lambda_\|^2(0)/\lambda_\|(T)^2$ calculated from the $\Delta\lambda_\|(T)$ data in Fig.~1, for several values of $\lambda_\|(0)$.  The thin lines are fits to the data with Eq.\ \protect\ref{dirtyd}. The inset shows the same data over a wider temperature range. }
\label{rfig}
\end{figure} 

In Fig.\ \ref{lfig} we show the temperature dependence of $\lambda_\|$ for a crystal of \etbr.  Our \ets~ samples show essentially the same $T$ dependence (see later).  The rapid variation of $\lambda_\|(T)$ at low temperature strongly suggests the existence of gapless excitations.  The similarities between these materials and the cuprates suggest that these excitations may originate from a $d$-wave pairing state.  In a pure $d$-wave superconductor it is well established that $\lambda_\|$ will vary linearly with $T$ at the lowest temperatures.  Although the data in Fig.\ \ref{lfig} show no region where $\lambda_\|(T)$ is strictly linear there are several factors which may intervene to mask this simple power law behavior.

First, Kosztin and Leggett  \cite{kosztin97} have argued that at low temperatures, of order $T_c~\xi_\|/\lambda_\|$, nonlocal effects can change the linear behavior to quadratic.  Both present materials however, are in the clean, local limit, $\lambda_\| \gg \ell \gg \xi_\|$ \cite{le92} and so this effect should only be important for $T \lesssim 70$ mK, which is well below our minimum temperature.  Next, there is the important distinction between $\Delta\lambda(T)$, which is directly measured, and the superfluid density [$\rho(T)= \lambda^2(0)/\lambda^2(T)$] which can only be inferred with knowledge of $\lambda(0)$.  In the $d$-wave model, the non-linear corrections to $\rhopar(T)$ at higher temperature are considerably smaller than for $\lambda_\|(T)$. If $\rhopar$ varies strictly linearly with $T$, $\rho=1-\alpha T/T_c$, then 
 $\lambda(T)=\lambda(0)[1+\frac{1}{2} \alpha (T/T_c) + \frac{3}{8}\alpha^2 (T/T_c)^2+\cdots$].  There is therefore invariably a quadratic component to $\lambda$ whose relative strength depends on $\alpha$ which in the $d$-wave model is inversely proportional to the angular slope of the energy gap at the nodes, $\alpha=\frac{4\ln2 T_c}{d\Delta(\theta)/d\theta|_{node}}$ \cite{xu95}.  A third factor is the influence of impurities.  Bonn {\it et al.} \cite{bonn94} have shown that in \ybco~ even small amounts of impurities can change the low temperature behavior of $\rho(T)$ from linear to quadratic.  Hirschfeld and Goldenfeld \cite{hirschfeld93} have explained how this behavior results from impurity scattering in a $d$-wave superconductor and have suggested the following formula to interpolate between the high temperature (pure) and the low temperature (impurity dominated) regimes
\begin{equation}
\Delta\rhopar(T)=\frac{\alpha T^2/T_c}{T+T^*}\quad,
\label{dirtyd}
\end{equation}
here $T^*$ parameterizes the impurity scattering rate.  For $T^*$ to be significant without $T_c$ being significant depressed requires that the scatterers be unitary \cite{hirschfeld93,bonn94}.  A residual small energy gap at the {\lq}node{\rq}, resulting from a mixed order parameter (e.g. $d_{x^2-y^2}+i\sqrt\epsilon s$) results in similar behavior \cite{xu95}.

In order to compare our data to the $d$-wave model we have plotted in Fig.\ \ref{rfig} the calculated superfluid density \cite{rhoscal}.   Muon spin relaxation measurements \cite{le92,uemura99} give values of $\lambda_\|(0)=0.78\pm0.2$~$\mu$m and $0.69\pm$~0.2$\mu$m for \etbr~and \ets~ respectively.  However, because of the very weak pinning and strong anisotropy of these compounds, extracting $\lambda(0)$ from $\mu$SR measurements can be problematic \cite{lee97} and therefore we show $\rhopar$ calculated for $\lambda_\|(0)$ in the range 0.8 to 1.8 $\mu$m.  The thin lines in the figure show a fit to the data with Eq.\ \ref{dirtyd}.   It can be seen that the fit is reasonable for all values of $\lambda_\|(0)$ but is best for the values close to the high end of our range.  For $\lambda(0)=1.8 \mu$m there is a quasilinear region extending from roughly 1~K to 6~K, and the fit to Eq.\ \ref{dirtyd} gives $\alpha=1.2$ and $T^*=0.6$~K.  Lower values of $\lambda_\|(0)$ results in a lower value of $T^*$ but restricts the upper limit of the quasilinear region; for $\lambda_\|(0)=0.8\mu$m, $T^*=0.4$~K and the curve breaks away from the linear fit at 1.7~K.  These values of $T^*/T_c$ are roughly comparable to what is found for good quality \ybco~ and other cuprates \cite{hardy93,bonn94,sflee96,broun97}.  If this low $T$ turnover is interpreted as arising from a mixed order parameter then our $T^*$ values would imply that the minimum gap is $\lesssim$ 3\% of the maximum \cite{xu95}.

For the lower values of $\lambda_\|(0)$ the shape of the $\rhopar(T)$ curve is unusual in that it has positive curvature.  Although we know of no cuprates which show this behavior it has been predicted in some $d$-wave models \cite{prohammer91}.  The shape of $\rhopar(T)$ for $\lambda_\|(0)\sim1.8\mu$m is similar to that reported for the single layer cuprate, Tl$_2$Ba$_2$CuO$_{6+\delta}$ \cite{broun97}. The values of $\alpha$ extracted from our fits are somewhat larger than that found for \ybco~ ($\alpha\sim 0.6$ \cite{hardy93,carrington99}) but are comparable to that found for Tl$_2$Ba$_2$CuO$_{6+\delta}$ \cite{broun97} ($\alpha\sim 1.0$).  This would imply that if \etbr~ is a $d$-wave superconductor then it has a small value of $d\Delta(\theta)/d\theta|_{node}$. We conclude therefore, that our data are compatible with a $d$-wave order parameter once the influence of impurities (or a small residual gap) and a large value of $\alpha$ are taken into account.  Although the above analysis has been conducted for \etbr, essentially the same conclusions are reached for \ets~ as $\Delta\lambda_\|(T)$ for the two compounds are very similar (see below).  The major difference is that if the $\rhopar(T)$ curves are forced to lie on top of those in Fig.\ \ref{rfig} then the $\lambda_\|(0)$ values for \ets~ are $\sim$35 \% lower (the lower limit of $\lambda_\|(0)$ in Fig.\ \ref{rfig} becomes $\sim$0.52$\mu$m).

\begin{figure}
\centerline{\psfig{figure=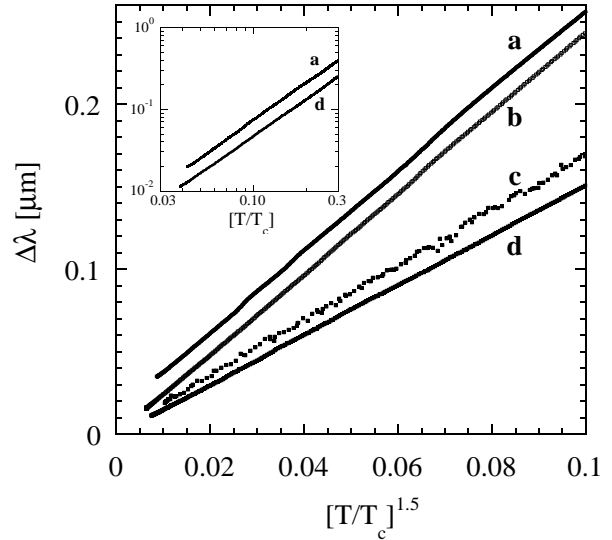,width=8cm}}
\vskip 0cm
\caption{$\Delta\lambda_\|(T)$ data for two samples of \etbr ~ ({\bf a, b}) and two samples of \ets~({\bf c, d}) plotted versus $[T/T_c]^\frac{3}{2}$.  The data have been offset for clarity.  The inset shows the data for samples {\bf a} and {\bf d} on a log scale.}
\label{T32fig}
\end{figure} 

We now consider an alternative interpretation of the data.  Kosztin {\it et al.} \cite{kosztin98} have recently proposed a BCS to Bose-Einstein  (BCS-BEC) crossover theory for short coherence length superconductors in which finite momentum pair excitations in addition to the usual fermionic Bogoliubov quasiparticles can deplete the condensate.  This theory has been used to explain the pseudogap behavior in the underdoped cuprates, and gives rise to a $T^\frac{3}{2}$ term in $\lambda$ below a characteristic temperature $T_{\rm BE}^*$.  NMR measurements indicate that pseudogap features are also present in the ET compounds \cite{kawamoto95} which leads us to consider this theory as a possible explanation for our data.

In Fig.\ \ref{T32fig} we have plotted our $\Delta\lambda_\|(T)$ data for two samples of  \etbr~and two samples of \ets~versus $T^\frac{3}{2}$.  It can been seen that the data follow this law very closely over almost a decade of temperature.    A fit of the low temperature data to a variable power law, $\Delta\lambda_\|(T)=\beta T^n$, gives $n=1.50\pm 0.04$, illustrating the robustness of the fractional exponent.  Within the errors the slope $\beta$ (not normalized to $T_c$) was the same for all samples ($\beta=0.06 \pm 0.01\mu$m K$^{-\frac{2}{3}}$).  The upper limit of the power law is $\sim 4$~K but is somewhat sample dependent.    

In the BCS-BEC theory both the magnitude of the $T^\frac{3}{2}$ term in $\lambda(T)$ and $T_{\rm BE}^*$ depend on material parameters. By modelling \etbr~ with a simple quasi-2D tight binding band structure and a moderately strong coupling constant, it can been shown \cite{kosztin98} that $\lambda_\|(T)$ follows a $T^\frac{3}{2}$ power law up to $\sim T_c/3$.  The crucial assumption of this fit is that the order parameter should have $s$-wave symmetry.  To get quantitative agreement requires that $\lambda_\|(0)\simeq3.2\mu$m for \etbr.  So although the theory explains well the power law displayed in Fig.\ \ref{T32fig}, quantitative agreement requires that $\lambda_\|(0)$ for our crystals is around 4 times the $\mu$SR estimates.

\begin{figure}
\centerline{\psfig{figure=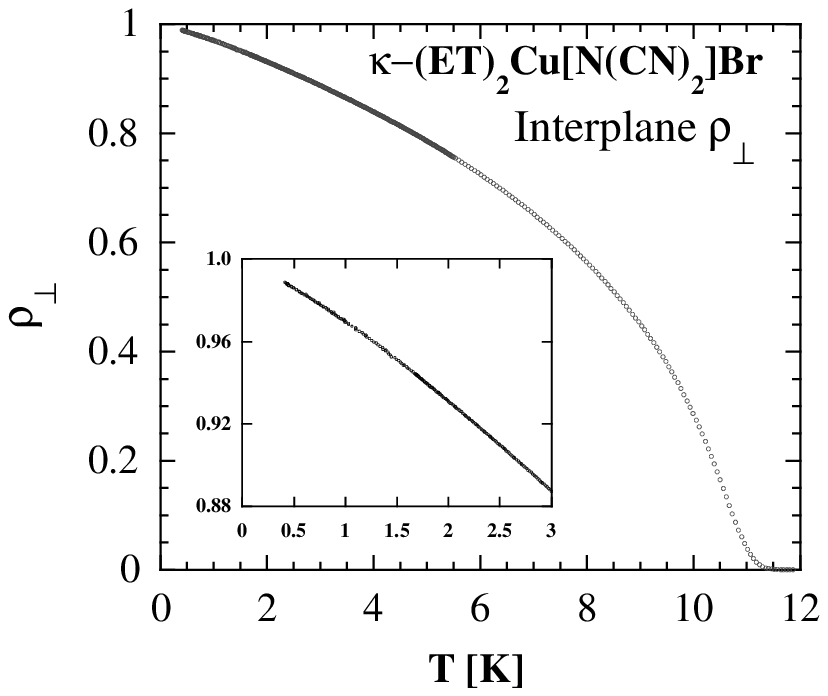,width=8cm}}
\vskip 0cm
\caption{The interplane superfluid density [$\rho_{\bot}=\lambda_\bot^2(0)/\lambda_\bot^2(T)$]  for a crystal of \etbr. The inset is an expanded view of the low temperature data.}
\label{lperpfig}
\end{figure}

Finally in Fig.\ \ref{lperpfig} we show data for the temperature dependence of the interplane penetration depth for a sample of \etbr.   As $\lambda_\bot(0)$ is large we are able to determine its absolute value directly from our measurements.  We obtain $\lambda_\bot(0)=100\pm20\mu$m which is comparable to previous measurements \cite{mansky94,shibauchi97,kirtley99}.  As shown in the figure, $\rho_\bot$ is also strongly temperature dependent at low temperature.  The relative strength of the temperature dependence of $\lambda$ in the two directions is $\Delta\lambda_\bot/\Delta\lambda_\| \simeq 17$ (between 0.4~K and 3~K).  $\rho_\bot(T)$ like $\rhopar(T)$ is close to being linear at low temperature.  A variable power law [$\rho_\bot(T)=1-\beta T^n$] fits the data well and gives an exponent $n=1.2\pm0.1$ (the uncertainty reflects the sample and $T$ fitting range dependence).  A dirty $d$-wave fit [Eq.\ (1)] is slightly worse and gives values of $T^*\simeq 0.5$~K and $\alpha=0.42\pm0.07$.

In these strongly anisotropic materials it has been argued that $\rho_\bot$ depends on both in-plane scattering processes as well as interplane hopping amplitudes and so is not in general expected to have the same $T$ dependence as $\rhopar$ \cite{radtke96}.  In HgBa$_2$CuO$_{6+\delta}$ it was found that although $\lambda_\| \sim T$, $\lambda_\bot\sim T^5$ \cite{panagopoulos97}, whereas in clean samples of Bi$_2$Sr$_2$CaCu$_2$O$_{8+\delta}$ both $\lambda_\|$ and $\lambda_\bot$ vary linearly with $T$ \cite{jacobs95}.  The fact that our data show a close to linear $T$ dependence however, supports our conclusion that these materials have low lying excitations.

In conclusion, our measurements give strong evidence for the existence of low lying excitations in both \etbr~ and \ets.   There are two competing explanations for our in-plane data.  We have shown that the observed non-linear $T$ dependence of $\lambda_\|$ can arise from a $d$-wave (or other gapless) order parameter with a small value of $d\Delta(\theta)/d\theta|_{node}$, and either a small residual gap ($\lesssim$ 3\% of the maximum) or a small density of unitary  impurity scatterers.  An alternative view is that these materials have a gapped order parameter and that the $T$ dependence of $\lambda_\|$ arises from incoherent pair excitations expected in short coherence length superconductors which exhibit a pseudogap \cite{kosztin98}.  Whereas the observed $T^\frac{3}{2}$ dependence of $\lambda_\|$ arises naturally in this model, the literature values of $\lambda_\|(0)$ are significantly closer to those required for the $d$-wave interpretation.  Resolution of this issue will require measurements of $\lambda_\|(T)$ to lower $T$ and more accurate determinations of $\lambda_\|(0)$.

We wish to acknowledge informative discussions with Q. Chen, K. Levin, P.J. Hirschfeld, J. Schmalian, R. Ramazashvili, N. Goldenfeld,  M.B. Salamon, Y.J. Uemura, S.L. Lee, A.J. Leggett and especially  I. Kosztin. This work was supported by Science and Technology Center for Superconductivity Grant No. NSF-DMR 91-20000.    Work at Argonne National Laboratory was sponsored by the U.S. Department of Energy, Office of Basic Energy Sciences, Division of Materials Sciences, under Contract W-31-109-ENG-38.


\begin{references}
\bibitem{averageinplane} Our in-plane measurements average over the two in-plane components to $\lambda$.
\bibitem{rhoscal} $\Delta\lambda(T)$ is defined to be zero at 0.4~K and, as shown below,  is found to vary as $T^\frac{3}{2}$ over a wide range.  In calculating $\rhopar$ we used this fit to estimate the small difference between  $\lambda(0)$ and $\lambda(0.4$~K).
\end{references}
\end{document}